\documentclass[aps,prd,preprint,superscriptaddress,tightenlines,nofootinbib]{revtex4}

\usepackage{graphicx}
\usepackage{dcolumn}
\usepackage{bm}
\usepackage[figuresleft]{rotating}

\newcommand{\dedx}{dE/dx}
\newcommand{\dele}{\Delta E}

\newcommand{\eff}{\epsilon}

\newcommand{\psipp}{\psi(3770)}

\newcommand{\EE}{e^+e^-}

\newcommand{\pip}{\pi^+}
\newcommand{\pim}{\pi^-}
\newcommand{\piz}{\pi^0}
\newcommand{\pipi}{\pip\pim}

\newcommand{\ra}{\rightarrow}

\newcommand{\ddb}{D\bar{D}}

\newcommand{\Dz}{D^{0}}

\newcommand{\Dp}{D^+}
\newcommand{\Dm}{D^-}
\newcommand{\enu}{e^+ \nu_e}

\newcommand{\mbc}{M_{\text{bc}}}
\newcommand{\ks}{K_S^0}

\newcommand{\etap}{\eta^\prime}

\newcommand{\GG}{\gamma\gamma}

\newcommand{\KK}{K^+K^-}

\newcommand{\bmath}{\begin{displaymath}}
\newcommand{\emath}{\end{displaymath}}

\newcommand{\beq}{\begin{equation}}
\newcommand{\eeq}{\end{equation}}
\newcommand{\bfg}{\begin{figure}}
\newcommand{\efg}{\end{figure}}
\newcommand{\bitm}{\begin{itemize}}
\newcommand{\eitm}{\end{itemize}}
\newcommand{\bnum}{\begin{enumerate}}
\newcommand{\enum}{\end{enumerate}}
\newcommand{\btbl}{\begin{table}}
\newcommand{\etbl}{\end{table}}
\newcommand{\btbu}{\begin{tabular}}
\newcommand{\etbu}{\end{tabular}}

\begin{document}

\preprint{CLNS 07/2010}       
\preprint{CLEO 07-14}         

\title{Observation of $D^+ \rightarrow \eta e^+ \nu_e $}

\author{R.~E.~Mitchell}
\author{M.~R.~Shepherd}
\affiliation{Indiana University, Bloomington, Indiana 47405 }
\author{D.~Besson}
\affiliation{University of Kansas, Lawrence, Kansas 66045}
\author{T.~K.~Pedlar}
\affiliation{Luther College, Decorah, Iowa 52101}
\author{D.~Cronin-Hennessy}
\author{K.~Y.~Gao}
\author{J.~Hietala}
\author{Y.~Kubota}
\author{T.~Klein}
\author{B.~W.~Lang}
\author{R.~Poling}
\author{A.~W.~Scott}
\author{A.~Smith}
\author{P.~Zweber}
\affiliation{University of Minnesota, Minneapolis, Minnesota 55455}
% ...
\collaboration{CLEO Collaboration}
\noaffiliation

\date{February 28, 2008}

\begin{abstract}
Using a 281 $\text{pb}^{-1}$ data sample collected at the $\psipp$
resonance with the CLEO-c detector at the Cornell Electron Storage
Ring, we report the first observation of $\Dp\ra\eta\enu$. We also
set upper limits for $\Dp\ra\eta'\enu$ and $\Dp\ra\phi\enu$ that
are about two orders of magnitude more restrictive than those
obtained by previous experiments.
\end{abstract}

\pacs{13.20.Fc}
\maketitle

The quark mixing parameters are fundamental constants of the
Standard Model (SM) of particle physics. They determine the nine
weak-current quark coupling elements of the
Cabibbo-Kobayashi-Maskawa (CKM) matrix~\cite{ckm}.
Charm semileptonic decays have been studied in considerable detail
because they provide direct measurements of the
magnitudes of the CKM elements $V_{cd}$ and $V_{cs}$,
and a stringent test of theoretical
predictions of strong
interaction effects in the decay amplitude.
In order to gain a complete understanding of charm semileptonic decays it is important to
study as many exclusive modes as possible.
Several rare modes have yet to be observed.

The most precise measurements to date of absolute branching
fractions of $D$-meson semileptonic decays have been made by
CLEO-c at the
$\psipp$~\cite{56pb_dp,56pb_d0,Nadia,Batbold}.
Absolute exclusive semileptonic branching fractions were measured for nine
final states that included a single $K$,
$\pi$, $K^*$, $\rho$, or $\omega$ meson. Summing the exclusive
semileptonic branching fractions gives
$\sum \mathcal{B}(D^0_{\rm excl}) = [6.3 \pm 0.2 {\rm (stat)} \pm 0.2 {\rm (syst)} ]$\% and
$\sum \mathcal{B}(D^+_{\rm excl}) = [15.2 \pm 0.3 {\rm (stat)} \pm 0.4 {\rm (syst)}]$\%.
CLEO has also measured the absolute branching
fractions of inclusive semileptonic decays and found
${\cal B}(\Dz\ra X\enu) = (6.46 \pm 0.17 \pm 0.13)$\% and
${\cal B}(\Dp \ra  X \enu)  = (16.13 \pm 0.20 \pm 0.33)$\%~\cite{281pbInc}.
These measurements are consistent with, but larger than, the sum
of the exclusive semileptonic branching
fractions. Although the possibility of
additional semileptonic modes of the $D^0$ and $D^+$ with large
branching fractions is excluded, a window for additional
semileptonic decay modes remains.

Semileptonic decays of a $D$ meson with an $\eta$ or $\etap$ in
the final state have not yet been observed. Their discovery would
open future opportunities to obtain information about $\eta -
\eta^\prime$ mixing~\cite{isgw2}. They also probe the composition of the $\eta$
and $\eta^\prime$ wave functions when combined with measurements
of the corresponding $D_s$ semileptonic decays~\cite{Bigi} and gauge the
possible role of weak annihilation in the corresponding
$D_s$-meson semileptonic decays~\cite{Bigi}. The process $D^+ \ra \phi e^+
\nu_e$ is not expected to occur in the absence of mixing between
the $\omega$ and $\phi$~\cite{pdg06}.

We report herein the first observation and absolute branching
fraction measurement of $\Dp\ra\eta\enu$, and results of searches
for $\Dp \to \etap\enu$ and $\Dp \to \phi\enu$. (Throughout this
Letter charge-conjugate modes are implied.) The data sample used
for these measurements consists of an integrated luminosity of
$281~\text{pb}^{-1}$ at the $\psipp$ resonance, and includes about
$8\times 10^5$ $\Dp\Dm$ events~\cite{dhad281}.
The data were
produced in $\EE$ collisions at the Cornell Electron Storage
Ring (CESR-c) and collected with the CLEO-c detector.
This is the same data set used in Refs.~\cite{Nadia,Batbold,281pbInc,dhad281}.

CLEO-c is a general-purpose solenoidal detector. The charged
particle tracking system covers a solid angle of 93\% of $4 \pi$
and consists of a small-radius six-layer low mass stereo wire
drift chamber concentric with, and surrounded by, a 47-layer
cylindrical drift chamber. The chambers operate in a 1.0 T
magnetic field and achieve a momentum resolution of $\sim$0.6\% at
$p=$1~GeV/$c$.  The main drift chamber provides
specific-ionization ($\dedx$) measurements that discriminate
between charged pions and kaons. Additional hadron identification
is provided by a Ring-Imaging Cherenkov (RICH) detector covering
approximately 80\% of $4 \pi$. Identification of positrons and
detection of neutral pions and eta mesons relies on an
electromagnetic calorimeter consisting of 7800
cesium iodide crystals and covering about 93\% of $4 \pi$. The
calorimeter achieves a photon energy resolution of 2.2\% at
$E_\gamma=$1~GeV and 5\% at 100~MeV. The CLEO-c detector is
described in detail elsewhere~\cite{cleo_detector}.

The technique for our analysis was first applied by the Mark III collaboration~\cite{mark3_tag} at SPEAR.
The presence of two $D^\pm$ mesons in a $D^+D^-$ event allows
a tag sample to be defined in which a $D^-$
is reconstructed in a hadronic decay mode.
A sub-sample is then defined in which a positron and a set of hadrons, as a signature
of a semileptonic decay, are required in addition to the tag.
The tag yield can be expressed
as $N_{{\rm tag} } = 2N_{DD}  { \cal B}_{{\rm tag}} \epsilon_{{\rm tag}}$, where $N_{DD}$ is the produced
number of $D^+D^-$ pairs, $ {\cal{B}}_{{\rm tag}}$ is the branching fraction of
hadronic modes used in the tag
sample, and $\epsilon_{{\rm tag}}$ is the tag efficiency.
The yield of tags with a semileptonic decay can be expressed as
$N_{{\rm tag, SL}} = 2N_{DD} {\cal{B}}_{{\rm tag}} {\cal{B}}_{{\rm SL}} \epsilon_{{\rm tag, SL}}$
where ${\cal{B}}_{{\rm SL}}$ is the semileptonic decay branching fraction,
including subsidiary branching fractions, and $\epsilon_{{\rm tag, SL}}$ is the efficiency of finding the tag and the semileptonic decay in the same event.
From the expressions for $N_{{\rm tag}}$ and  $N_{{\rm tag, SL}}$
we obtain
\begin{equation}
{\cal{B}}_{{\rm SL}} =  \frac {N_{ \rm tag, SL }} {N_{ \rm tag}}   \frac {\epsilon_{ \rm tag}} {\epsilon_{\rm tag, SL}} =
\frac {N_{ \rm tag, SL} / \epsilon} {N_{ \rm tag}},
\label{eq:master}
\end{equation}
where $\epsilon = \epsilon_{ \rm tag, SL}/\epsilon_{ \rm tag }$
is the effective signal efficiency.
The branching fraction determined by tagging is an absolute
measurement. It is independent of the integrated luminosity and
number of $\Dp$ mesons in the data sample.
Due to the large solid angle acceptance
and high segmentation of the CLEO-c detector and the low multiplicity of the events
$\epsilon_{ \rm tag, SL} \approx \epsilon_{\rm tag} \epsilon_{\rm SL}$, where $\epsilon_{ \rm SL}$ is the semileptonic decay efficiency. Hence the ratio $  \epsilon_{\rm tag, SL }/\epsilon_{\rm tag}$
is insensitive to most systematic effects associated with the tag mode and the
absolute branching fraction determined with this procedure is nearly independent of the tag mode.

Candidate events are selected by reconstructing a $\Dm$ tag
in one of the following six hadronic final states:
$\ks\pim$, $K^+ \pim\pim$, $\ks \pim\piz$, $K^+ \pim\pim\piz$, $\ks \pim\pim\pip$, and $\KK\pim$. Tagged events are selected based on two variables: $\dele \equiv
E_D - E_{\text{beam}}$, the difference between the energy of the
$\Dm$ tag candidate ($E_D$) and the beam energy
($E_{\text{beam}}$), and the beam-constrained mass $\mbc \equiv
\sqrt{E^2_{\text{beam}}/c^4 - |\vec{p}_D|^2/c^2}$, where
$\vec{p}_D$ is the measured momentum of the $\Dm$ candidate.
Selection criteria for tracks, $\piz$ and $\ks$ candidates used in
the reconstruction of tags are described in Ref.~\cite{56pb_had}.
If multiple candidates are present in the same tag mode, one
candidate per tag charge with the smallest $\Delta E$ is chosen.
The yield of each tag mode, and the combined yield of all six tag
modes, are obtained from fits to the $\mbc$ distributions,
where the signal shape
includes the effects of beam energy smearing, initial
state radiation, the line shape of the $\psi(3770)$, and
reconstruction resolution, and the background is described by an ARGUS
function~\cite{ARGUS}, which models combinatorial contributions.
The
data sample comprises approximately 163,000 reconstructed charged
tags (Table~\ref{tag_yield}).

\begin{table}[ht]
\caption{Tag yields of the six $\Dm$ hadronic modes with statistical uncertainties.
 }
\begin{center}
\begin{tabular}{lcrr}\hline\hline
Tag Mode   & \multicolumn{3}{c}{$N_{\text{tag}}$} \\\hline
$\Dm \ra K^0_S \pim$         &11469&$\pm$& 116  \\
$\Dm \ra K^+ \pim \pim$      &79933&$\pm$& 293  \\
$\Dm \ra K^0_S \pim \piz$    &25243&$\pm$& 212  \\
$\Dm \ra K^+ \pim\pim\piz$   &23733&$\pm$& 185  \\
$\Dm \ra K^0_S \pim\pim\pip$ &16446&$\pm$& 177  \\
$\Dm \ra K^- K^+ \pim$       &6785 &$\pm$& 97 \\
\hline
All Tags                   & 163057&$\pm$& 483 \\
\hline\hline
\end{tabular}
\end{center}
\label{tag_yield}
\end{table}

After a tag is identified, we search for a positron and a set of
hadrons recoiling against the tag. (Muons are not used because the
CLEO-c muon identification system has poor acceptance in the
momentum range characteristic of semileptonic $D$ decays at the
$\psipp$.)
Positron candidates are selected based on a likelihood ratio constructed
from three inputs: the ratio of the energy deposited in the calorimeter
to the measured momentum ($E/p$), $dE/dx$, and RICH information~\cite{56pb_d0}.
Furthermore, candidates must have
momenta of at least
200~MeV/$c$ and satisfy $|\cos\theta|<0.90$, where $\theta$ is
the angle between the positron direction and the beam axis. The
efficiency for positron identification has been measured primarily
with radiative Bhabha events.
In the kinematic region used in this analysis,
the efficiency
rises from about 50\% at 200~MeV/$c$ to 95\% just
above 300~MeV/$c$ and is roughly constant thereafter. The rates
for misidentifying charged pions and kaons as positrons, averaged
over the momentum range, is approximately 0.1\%. Bremsstrahlung
photons are recovered by adding showers within $5^\circ$ of the
positron that are not matched to other particles.

Hadronic tracks produced in semileptonic $\Dp$ decay are required
to have momenta above 50~MeV/$c$ and $|\cos\theta|<0.93$. Pion and
kaon candidates are required to have $\dedx$ measurements within
three standard deviations ($3\sigma$) of the expected value. For
tracks with momenta greater than 700~MeV/$c$, RICH information, if
available, is combined with $\dedx$. The efficiencies (95\% or
higher) and misidentification rates (a few per cent) are
determined with charged pion and kaon samples from hadronic $D$
decays.

We select $\piz$ candidates from pairs of photons, each having an
energy of at least 30~MeV, and a shower shape consistent with that
expected for a photon.  A kinematic fit is performed constraining
the invariant mass of the photon pair to the  known $\piz$ mass.
The candidate is accepted if the unconstrained invariant mass is
within 3$\sigma$ of the nominal $\piz$ mass, where $\sigma$
(typically 6 MeV/$c^2$) is determined for that candidate from the
kinematic fit, and the kinematic parameters for the $\piz$
determined with the fit are used in further reconstruction.
We reconstruct $\eta$ candidates in two decay modes. For the decay
$\eta\ra\GG$, candidates are formed using the same procedure as
for $\piz$ except that $\sigma\sim$ 12~MeV/$c^2$. For
$\eta\ra\pipi\piz$ we require that the invariant mass of the three
pions be within 12~MeV/$c^2$ of the known $\eta$ mass. We
reconstruct $\etap$ in the decay mode $\etap \to \pipi \eta $. We
require $|m_{\pipi\eta} - m_{\etap}|< 10$~MeV/$c^2$.
We reconstruct $\phi$ candidates in the mode $\phi\ra\KK$
requiring $|m_{K K}-m_{\phi}|<13.5$~MeV/$c^2$. For both the
$\etap$ and $\phi$ these mass cuts correspond to $\pm 3 \sigma$.

The $D^-$ tag and $D^+$ semileptonic decay are combined if
they account for all tracks in the event.
Semileptonic decays are identified with
$U \equiv E_{\text{miss}} - c|\vec{p}_{\text{miss}}|$, where
$E_{\text{miss}}$ and $\vec{p}_{\text{miss}}$ are the missing energy and
momentum of the $D^+$ meson. If the decay products have been correctly identified,
$U$ is expected to be zero, since only a neutrino is undetected.
The resolution in $U$ is improved
by constraining the magnitude and direction of the $D^+$ momentum to be
$p_{D^+} =\sqrt{E^2_{\rm beam}/c^2-c^2m^2_{D}}$, and
$\vec{p}_{D^+} = - \vec{p}_{D^-}$~\cite{56pb_d0}, respectively.
Due to the finite resolution of the detector,
the distribution in $U$ is approximately Gaussian, centered at $U
= 0$ with $\sigma\sim 10$~MeV (the width varies by mode).
The number of events with multiple candidates varies by mode,
ranging from zero for $D^+ \to  \phi \enu$ to 83\%
for $D^+ \to \eta^\prime (\pi \pi \eta, \eta  \to \pi
\pi \pi^0) \enu$. To
remove multiple candidates in each semileptonic mode one
combination is chosen per tag mode per tag charge, based on the
proximity of the invariant masses of the $\piz$, $\eta$, $\etap$,
or $\phi$ candidates to their expected masses.

\begin{figure}
\includegraphics*[width=3.2in]{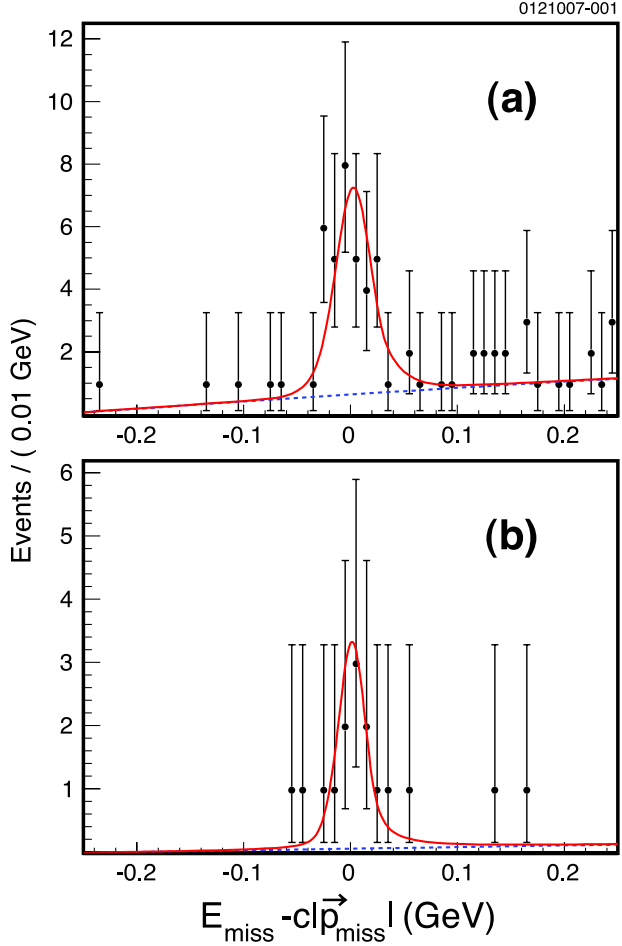}
\caption{Fits to the $U$ distributions in data (filled circles
with error bars) for $\Dp\ra\eta\enu$ for two $\eta$ decay modes:
(a) $\eta\ra\GG$, (b) $\eta\ra\pipi\piz$. The solid line
represents the fit of the sum of the signal function and
background function to the data. The dashed line indicates the
background contribution. } \label{fig:etaU}
\end{figure}

The $U$ distributions for $\Dp\ra\eta\enu$ for each $\eta$ decay
mode with all tag modes
combined are shown in Fig.~\ref{fig:etaU}.
The $U$ distributions in data for
$\Dp\ra\etap(\pipi\eta)\enu$ and
$\Dp\ra\phi(\KK)\enu$ have no entries in  $-250  < U <
250$ MeV and are not included in Fig. 1~\cite{extra-mode}. The
yield for $\Dp\ra\eta\enu$ is determined
from a binned likelihood
fit to the $U$ distribution where the signal is
described by a modified Crystal Ball function with two power-law
tails~\cite{cb_2tail} which account for initial- and final-state
radiation (FSR) and mismeasured tracks. The signal parameters
are fixed with a GEANT-based Monte Carlo (MC)
simulation~\cite{geant} in fits to the data. The background functions are determined
by a simulation that incorporates all available data on $D$ meson
decays.  The backgrounds are small and arise mostly from
misreconstructed semileptonic decays with correctly reconstructed
tags. For $\Dp\ra\eta(\GG)\enu$, there is background
from $\Dp\ra\ks (\piz\piz) \enu$ and $\Dp\ra\piz\enu$. The background shape parameters are fixed, while the background normalizations are allowed
to float in fits to the data. The signal yields $N_{\rm tag, SL}$ are given in Table~\ref{br}.
The fits describe the data well.
After increasing the backgrounds by one statistical $\sigma$,
the probabilities that background fluctuations account for the signals
are $2 \times 10^{-10}$ and $6 \times 10^{-8}$, corresponding to
$6.3 \sigma$  and $5.3 \sigma$, for
$D^+ \to \eta(\gamma\gamma)\enu$ and $D^+ \to \eta(\pi^+\pi^-\pi^0)\enu$,
respectively.
This is the first
observation of $\Dp\ra\eta\enu$.

The absolute branching fractions and 90\% confidence level (C.L.) upper limits
in Table~\ref{br} are obtained from
Eq.~(\ref{eq:master}). The signal efficiencies, $\epsilon$, are determined by MC simulation, and have been weighted
by the tag yields shown in Table~\ref{tag_yield}.
Due to differences between the simulation and data, corrections
are applied to $N_{\rm{tag, SL}}$
for $\pi^0$ and $\eta$ finding, positron identification, and
charged $\pi$ and $K$ identification.
The corrections are given in Table~\ref{totalsys}.


\begin{sidewaystable*}[tb]
\caption{Signal efficiencies, yields
(90\% C.L. intervals),
and
branching fractions (90 \% C.L. upper limits)
for $\Dp\ra\etap\enu$, ($\Dp\ra\etap\enu$ and $\phi\enu$)
in units of $\times 10^{-4}$ in this work and
comparisons to PDG~\cite{pdg06} and two theoretical
predictions ISGWII~\cite{isgw2} and FK~\cite{fajfer}.
In the fourth column the first
uncertainty is statistical and the second systematic. In other
columns the uncertainty is statistical or statistical and
systematic combined in quadrature.
The efficiencies include subsidiary branching fractions
from PDG~\cite{pdg06}.
}
\begin{center}
\begin{tabular}{lccccccc}\hline\hline
Decay Mode   &$\eff$ (\%) &  $N_{\text{tag, SL}} $ & ${\cal B}_{\rm SL} $ &
$~{\cal B}_{\rm SL}$(PDG)& $~{\cal B}_{\rm SL} $(ISGWII) & $~{ \cal B}_{\rm SL}$(FK)
\rule[-1mm]{0mm}{4.3mm}   \\
\hline
$\Dp\ra\eta(\GG)\enu$       &~$15.13\pm0.07$~ & ~~~$32.6\pm6.7$~~~ &$13.2\pm2.3\pm0.6$ &       &     &   \rule[-1mm]{0mm}{4.3mm}    \\
$\Dp\ra\eta(\pipi\piz)\enu$ &~~$5.99\pm0.04$ & $13.3\pm4.0$ &$13.6\pm3.7\pm0.5$ &       &     &    \\
$\Dp\ra\eta\enu$ (Combined)   &              &              &$13.3\pm2.0\pm0.6$ & $<$70 & 11  & 10 \\\hline
$\Dp\ra\etap(\pipi\eta\ra\pipi\GG)\enu$    &~~$3.38\pm0.02$ & (0.00, 2.30) & $<$4.4&&&\rule[-1mm]{0mm}{4.3mm}   \\
$\Dp\ra\etap(\pipi\eta\ra 2(\pipi)\piz)\enu$&~~$0.85\pm0.01$ & (0.00, 2.30) & $<$17.3&&&\\
$\Dp\ra\etap\enu$ (Combined)                 &~~$4.23\pm0.02$ &
(0.00, 2.30) & $<$3.5 & $<$110   & 5 & 1.6\\\hline
$\Dp\ra\phi(\KK)\enu$                      &~~$8.97\pm0.10$ &
(0.00, 2.30) & $<$1.6 & $<$209   &   &    \rule[-1mm]{0mm}{4.3mm}   \\\hline \hline
\end{tabular}
\end{center}
\label{br}
\end{sidewaystable*}

\begin{table}[tbp]
\caption{Summary of corrections and systematic uncertainties
to ${\cal B}_{\rm SL}$ in percent (\%).
A minus sign indicates the MC simulation has higher
efficiency than the data.}
\begin{center}
\begin{tabular}{lcc}\hline\hline
Source                                &Correction (\%)   & Uncertainty (\%)
\\\hline
Tracking                              & -           & 0.3 - 1.5 \\
$\eta\ra\gamma\gamma$ reconstruction  &$-$6.5       &       4.0 \\
$\piz$ finding                        &$-$5.8 - $-$5.0& 2.5 - 3.6\\
Electron identification               & $-$1.2 - $-$1.0  &1.0 \\
Background shape                      &  - & 0.0 - 1.8 \\
Hadron identification                 & $-$6.9 - $-$2.2     &0.2 - 0.8 \\
Number of $D$ tags                    & -              &0.5  \\
Veto of unused tracks                 & - &0.3  \\
Signal Shape                          & - &0.0 - 0.4  \\
Simulation of form factors            & - & 1.0 - 3.0  \\
Simulation of FSR                     & - & 0.6 \\
MC statistics                         & - & 0.9 - 1.8 \\ \hline
Total                                 & $-$12.6 - $-$7.6  & 3.8 - 4.8 \\
\hline\hline
\end{tabular}
\end{center}
\label{totalsys}
\end{table}

We consider the following sources of systematic uncertainty and
give our estimates of their magnitudes in Table~\ref{totalsys}.
Where these estimates vary by decay mode a range is given.
The uncertainty in the track finding efficiency
is estimated with missing mass
techniques applied to the data and simulation~\cite{dhad281}.
The correction to the $\piz \ra \GG$ $(\eta\ra\GG)$ detection
efficiency and its uncertainty are estimated with
the same method.
The momentum-dependent correction to the positron identification, and its uncertainty,
are obtained from comparisons of the detector response to
positrons from radiative Bhabhas in data and MC simulations~\cite{56pb_d0}.
The corrections to the charged pion and kaon identification
efficiencies, and the associated uncertainties,  are
estimated using hadronic $D$-meson decays.
The uncertainty in the number of tags is estimated by using
alternative signal functions in the fits, and by counting.
The uncertainty in modelling the background shapes in the fits to
the $U$ distributions has contributions from the uncertainties in
the simulation of the positron and hadron fake rates.
The uncertainty associated with the requirement that there be no
additional tracks in tagged semileptonic events is estimated by
comparing fully reconstructed $\ddb$ events in data and
simulation.
The uncertainty associated with the shape of the signal function
is estimated by using alternative signal functions.
The uncertainty in the semileptonic reconstruction efficiencies
due to imperfect knowledge of the semileptonic form factors is
estimated by varying the form factors in the corresponding
well-studied transitions $D \ra K/ \pi /K^*
\enu$, by their uncertainties and assuming
that the resulting changes in efficiency are a measure of the
uncertainty in efficiencies for the semileptonic transitions being
studied here~\cite{56pb_dp}.
The uncertainty associated with the simulation of FSR and
bremsstrahlung in the detector material is estimated by varying
the amount of FSR modelled by the PHOTOS algorithm~\cite{photos}
and by repeating the analysis with and without recovery of FSR
photons.
The uncertainty associated with the simulation of initial-state
radiation ($\EE\ra\ddb \gamma$) is negligible.
These estimates are added in quadrature
to obtain the total systematic uncertainties reported in
Table~\ref{br}.

For the upper limits in Table~\ref{br}, systematic uncertainties
are incorporated by combining
with statistical uncertainties in quadrature and increasing the upper
limit on the number of observed events by one $\sigma$
of the combined uncertainty. Systematic uncertainties for all
modes are much smaller than statistical uncertainties.

Our branching fractions for $\Dp\ra\eta\enu$ and upper limits for
$\Dp\ra\etap\enu$ and $\Dp\ra\phi\enu$ are compared to previous
upper limits~\cite{pdg06} in Table.~\ref{br}. The branching fractions measured
using the two $\eta$ decay modes are consistent, and
the combined branching fraction is consistent with the PDG upper
limit.
Theoretical predictions
for ${\cal B} (\Dp\ra\eta\enu)$ and ${\cal B} (\Dp\ra\etap\enu)$
in the ISGW2 model~\cite{isgw2} and a model (FK) which combines
heavy-quark symmetry and properties of the chiral
Lagrangian~\cite{fajfer}, are also listed in Table.~\ref{br}. Our
${\cal B}(\Dp\ra\eta\enu)$ is consistent with these predictions,
although our statistical uncertainty is large.
The upper limits for $\Dp\ra\eta'\enu$ and $\Dp\ra\phi\enu$ are
about two orders of magnitude more restrictive than previous
experimental limits.

In summary, we have made the first observation of $\Dp\ra\eta\enu$
and measured the branching fraction, which is found to be
consistent with model predictions.  The sum of exclusive semileptonic
branching fractions of the $D^+$ is approximately $0.9\%$
smaller than the measured inclusive semileptonic branching fraction.
Our ${\cal B}(\Dp\ra\eta\enu)$ increases the exclusive sum by about 0.1\%,
indicating that further exclusive rare semileptonic modes may await discovery.
We have searched for the
decays $\Dp\ra\etap\enu$ and $\Dp\ra\phi\enu$ and set significantly improved upper limits for each mode. The predictions for ${\cal B
}(\Dp\ra\etap\enu)$~\cite{isgw2,fajfer} are similar in magnitude
to our upper limit. If these models are correct, an observation of
$\Dp\ra\etap\enu$ is likely in the near future.

We gratefully acknowledge the effort of the CESR staff
in providing us with excellent luminosity and running conditions.
This work was supported by
the A.P.~Sloan Foundation,
the National Science Foundation,
the U.S. Department of Energy, and
the Natural Sciences and Engineering Research Council of Canada.


\begin{thebibliography}{99}

\bibitem{ckm} M.~Kobayashi and T.~Maskawa, Theor. Phys. {\bf 49}, 652 (1973).
\bibitem{56pb_dp}  G.S.~Huang~{\it et~al.} [CLEO Collaboration],
     Phys. Rev. Lett. {\bf 95}, 181801~(2005).
\bibitem{56pb_d0}      T.E.~Coan {\it et~al.} [CLEO Collaboration],
     Phys. Rev. Lett. {\bf 95}, 181802~(2005).
\bibitem{Nadia}
D. Cronin-Hennessey~{\it et al.}, [CLEO Collaboration],
     Phys. Rev. Lett. {\bf 100}, 251802 (2008), and
S.~Dobbs~{\it et al.}, [CLEO Collaboration], Phys. Rev. D {\bf 77}, 112005 (2008).
\bibitem{Batbold} J. Y. Ge~{\it et al.}, [CLEO Collaboration],
  arXiv:0810.3878 submitted to Phys, Rev. D (2008).
\bibitem{281pbInc}  N.E.~Adam~{\it et~al.} [CLEO Collaboration],
     Phys. Rev. Lett. {\bf 97}, 251801~(2006).
\bibitem{isgw2} D.~Scora and N.~Isgur, Phys. Rev. D {\bf 52}, 2783 (1995).
\bibitem{Bigi} S. Bianco, F. Fabbri, D. Benson, and I. Bigi, La Rivista del Nuovo Cimento, 26, {\bf 7-8} (2003).
\bibitem{pdg06}  W.-M.~Yao {\it et al.} [Particle Data Group], J. Phys. G {\bf 33}, 1 (2006).
\bibitem{dhad281}
  S.~Dobbs {\it et al.}  [CLEO Collaboration],
  Phys.\ Rev.\  D {\bf 76}, 112001 (2007).

\bibitem{cleo_detector} 
Y.~Kubota {\it et~al.}, {Nucl. Instrum. Meth. Phys. Res., Sect. A} \textbf{320}, {66} ({1992});
D.~Peterson {\it et~al.}, {Nucl. Instrum. Meth. Phys. Res., Sect. A} \textbf{478}, {142} ({2002});
M.~Artuso {\it et~al.}, {Nucl. Instrum. Meth. Phys. Res., Sect. A} \textbf{554}, {147} ({2005}).

\bibitem{mark3_tag}  J.~Adler {\it et~al.} [Mark III Collaboration], Phys. Rev. Lett. {\bf 62}, 1821~(1989).

\bibitem{56pb_had}  Q.~He~{\it et~al.} [CLEO Collaboration], Phys. Rev. Lett. {\bf 95}, 121801~(2005).

\bibitem{ARGUS} H.~Albrecht {\it et al.}, [ARGUS Collaboration],
  Phys. Lett. B {\bf 241}, 278 (1990).

\bibitem{extra-mode} We have also searched for $\Dp\ra\etap\enu$
with $\etap \ra \rho^0\gamma$. This mode has a large branching
fraction and detection efficiency but also large background. No
significant signal was observed. Uncertainties in the simulation
of the background complicate extraction of an upper limit and so
this mode is not used.

\bibitem{cb_2tail}
T.~Skwarnicki, Ph.D thesis, Jagiellonian University in Krakow, 1986, DESY Report No. F31-86-02.

\bibitem{geant} R.~Brun {\it et~al.}, {\tt GEANT 3.21}, CERN Program Library Long Writeup W5013, unpublished.

\bibitem{photos} E.~Barberio and Z.~Was, Comput. Phys. Commun. {\bf 79}, 291 (1994).





\bibitem{fajfer} S.~Fajfer and J.~Kamenik, Phys. Rev. D {\bf 71}, 014020 (2005).

\end{thebibliography}
\end{document}